\documentclass[twocolumn,twoside,slac_two]{revtex4}
\usepackage{graphicx}
\usepackage{fancyhdr}
\begin{document}

\title{CDF Hot Topics}

%

\author{S. Donati}
\affiliation{University of Pisa, Pisa, Italy}

\begin{abstract}
In this paper we review the most recent CDF results 
in the fields of $b$ and $c$ Physics. 
\end{abstract}

\maketitle

\thispagestyle{fancy}


\section{Introduction}
B hadrons are abundantly produced at the Tevatron Collider, 
the measured B$^+$ cross section is 2.78$\pm$0.24~$\mu$b
in the region of transverse momentum $p_T(B^+) >$~6.0~GeV/c 
and rapidity $\mid y(B^+)\mid <$~1~\cite{Bcrosssection}. 
This cross section is three orders of magnitude larger than 
at $e^+ e^-$ machines
running at the $\Upsilon (4S)$ and the available energy allows 
the production of the heavier $B^0_s$, $B_c$ and $\Lambda_b$ hadrons. 
The challenge is extracting the interesting
B signals from a level of background which is three orders of
magnitude higher at production. This is achieved at CDF~II with 
dedicated detectors and triggers. The current CDF hot topics
reviewed in this paper are the recent analysis which sets the 
first bound on the mixing induced CP violation in the 
$B^0_s\rightarrow J/\psi \phi$ decays, the analysis leading
to the world best limits on the 
$B^0_s\rightarrow\mu^+\mu^-$ branching fraction, the $B_c$
and $B^0_s$ lifetime measurements, the analysis leading to
the evidence for the $D^0$ mixing and the search for the
$D^0\rightarrow\mu^+\mu^-$ decay.

\section{The Tevatron Collider and the CDF~II Detector}
The Tevatron Collider collides 36 $p\overline{p}$ bunches
at $\sqrt{s}$ = 1.96 TeV.  The design instantaneous luminosity
was 10$^{32}$ cm$^{-2}$s$^{-1}$ but the Tevatron largely 
exceeded it and set the peak luminosity record above 
$3.0\times 10^{32}$ cm$^{-2}$s$^{-1}$. With an already
integrated luminosity larger than 3.2~fb$^{-1}$, the
expectation is to have integrated $\sim$6~fb$^{-1}$ 
by the year 2010.

\subsection{Tracking Detectors}
The CDF~II tracker is located within a 14.1 kG solenoidal magnetic
field and it is composed of silicon detectors and a drift
chamber.
There are three independent silicon detectors, SVXII, ISL and
L00, for a total of eight silicon layers, 704 ladders and 722,432 
channels~\cite{svxii}.
SVXII is made of 360 double-sided ladders in a layout of six 15 cm
axial sections $\times$ twelve 30$^{\circ}$ $\phi$ slices 
$\times$ five radial layers between 2.5 and 10.6 cm from the beamline.
ISL covers the area between SVXII and the drift chamber,
with 296 double-sided ladders at radii of 20 and 28 cm. 
With a length of 1.9 m, it provides silicon hits out to $\mid \eta\mid <$~2.
L00 is a single-sided layer of 48 ladders mounted directly on
the beampipe, 1.5 cm from the beamline, which enhances the 
track impact parameter resolution.
The three subdetectors share the same readout system, starting
with the SVX3D chip, a custom designed ASIC with a 128 channel
$\times$ 42 capacitor analog storage ring, which makes it possible 
to acquire data in deadtimeless mode, integrating charge on one
capacitor while reading out another one. The data 
acquisition system provides silicon data in time for Level 2
trigger processing. Over 90~\% of the silicon detector is 
powered and more than 80~\% is providing quality data.
Charge collection efficiency is ~99~\% , with a single
hit efficiency~$>$~90~\%. The hit resolution for a two-strip
cluster is 9~$\mu$m. The signal-to-noise ratio is above
10:1 both for $r-\phi$ strips and for $r-z$ strips.
The Central Outer Chamber (COT, \cite{cot}) is located outside 
the silicon detectors and inside the time-of-flight detector
scintillators. The active volume of the COT spans 310~cm
in the beam direction, 43.4~cm and 132.3~cm in radius,
and the entire azimuth. The COT contains 30,240 sense wires
that run the length of the chamber between two end plates.
Approximately half of the wires are axial (run along the
$z$ direction) and half are small angle (2$^{\circ}$) stereo.
The $r-\phi$ view provides information for the $p_T$ measurement,
the $r-z$ view for the $\eta$ measurement.
The COT contains 96 sense wire layers in radius
that are grouped into eight superlayers. Each superlayer
is divided into supercells along the azimuthal angle, and
each supercell has 12 sense wires and a maximum drift distance 
that is approximately the same for all superlayers. Therefore
the number of supercells in a given superlayer scales
approximately with the radius of the superlayer. The 
supercell layout consists of a wire plane containing
sense and potential (or field shaping) wires and a field
(or cathode) sheet on either side. Each field sheet is
shared with the neighboring cell. The supercell is 
tilted by 35$^{\circ}$ with respect to the radial direction 
to compensate for the Lorentz angle of the drifting electrons
in the magnetic field. The gas is a mixture or Ar/Et (50:50)
and Isopropyl which provides a maximum drift time of 177~ns
on the maximum drift distance of 0.88~cm. The measurement
of the pulse widths provides a measurement of the dE/dx
in the chamber, used for particle identification.
The achieved performance of the integrated CDF~II tracker
is a transverse momentum 
resolution $\sigma (p_T)/p_T^2$ = 0.15~\% (GeV/c)$^{-1}$
and an impact parameter resolution
$\sigma (d)$ = 35 $\mu$m @2~GeV/c. This performance is
sufficient for the B physics analyses.

\subsection{Particle Identification Detectors}
CDF~II uses two detectors and two complementary techniques
for particle identification, one is the dE/dx measurement 
in the COT, the other one is the time-of-flight measurement
in a dedicated detector.
The COT readout electronics allows to measure 
the pulse width, which is related to the amount of charge 
collected by the wire. The truncated mean (80~\%) computed on 
the hits
associated to a track provides a measurement of the specific
ionisation (dE/dx) in the chamber. A detailed calibration of
the dE/dx measurement has been performed using 
samples of kaons and pions 
from $D^{*+}\rightarrow D^0 \pi^+\rightarrow [K^-\pi^+]\pi^+$,
protons from $\Lambda^0\rightarrow p \pi^-$, and muons
and electrons from $J/\psi\rightarrow\mu^+\mu^-$ and
$J/\psi\rightarrow e^+ e^-$. The achieved $K/\pi$ separation
for $p_T>$~2~GeV/c is 1.4$\sigma$.

The Time-of-Flight detector (TOF, \cite{tof}) is installed between
the drift chamber and the solenoid magnet and extends 4.7~cm
radially at a radius of roughly 138~cm. The detector is composed
by Bicron scintillator bars BC-408, selected for the long (2.5~m)
attenuation length. The bars have a dimension of 4$\times$4~cm$^2$
in cross-section and 279~cm in length. There are a total of 216
bars, each covering 1.7$^{\circ}$ in $\phi$ and $\mid\eta\mid<$1.
The photomultipliers are attached to each end of every bar.
The time resolution on the single hit is 110~ps and the $K/\pi$
separation is better than 2$\sigma$ for $p_T <$~1.5~GeV/c.
By combining the dE/dx and the time-of-flight measurements,
the achieved $K/\pi$ separation is better than 1.4~$\sigma$ 
in the entire momentum range.

\subsection{Lepton Detectors}
Segmented electromagnetic and hadronic calorimeters surround
the tracking system~\cite{cem}.
The electron energy is measured by lead-scintillator
sampling calorimeters. In the central region ($\mid\eta\mid<$1.1)
the calorimeters
are arranged in a projective barrel geometry and measure 
electromagnetic energy with a resolution of
$\mid \sigma(E_T)/E_T\mid^2 = (13.5~\%)^2/E_T(GeV)+(2~\%)^2$.
In the forward region (1.2$<\mid\eta\mid<$3.5) the calorimeters
are arranged in a projective end-plug geometry and measure
the electromagnetic energy with a resolution of
$\mid \sigma(E_T)/E_T\mid^2 = (14.4~\%)^2/E_T(GeV)+(0.7~\%)^2$.
Both central and forward electromagnetic calorimeters are
instrumented with finely segmented detectors which measure
the shower position at a depth where the energy deposition 
by a typical shower reaches its maximum.
The central muon detector (CMU \cite{cmu}) is located around 
the outside of the central hadron calorimeter at a radius of 
347~cm from the beam axis.
The calorimeter has a thickness of 5.5 interaction lengths
and a $\phi$ segmentation of 15$^{\circ}$. The muon drift 
cells are 226~cm
long and cover 12.6$^{\circ}$ in $\phi$, giving a $\phi$
coverage of 84~\%. The pseudorapidity coverage is $\mid\eta\mid <$~1.
Each module consists of four layers of four rectangular drift
cells. The sense wires in alternating layers are offset by
2~mm for ambiguity resolution. The smallest unit in the CMU,
called a stack, covers about 1.2$^{\circ}$ and includes four
drift cells, one from each layer. Adjacent pairs of stacks
are combined together to form a two-stack unit called a tower.
A track segment detected in these chambers is called a CMU
stub. A second set of muon chambers is located behind an
additional 60~cm of steel. The chambers are arranged axially
to form a box around the central detector. The coverage of the
central muon system is extended to the region 0.6$< \mid\eta\mid <$1.0
by four free-standing conical arches which hold drift chambers
which cover 71~\% of the solid angle.

\subsection{Trigger and Data Acquisition}
CDF~II uses a three-level system to reduce the 1.7~MHz bunch 
crossing rate to 100~Hz written on tape. The Level 1 is a 
deadtimeless 7.6 MHz synchronous pipeline with 42 cells, 
which allows 5.5~$\mu$s to form a trigger decision. The
maximum sustainable Level 1 output rate is $\approx$30~kHz.
The Level 2 is an asynchronous pipeline with an average
latency of 20~$\mu$s. While the events accepted by Level~1
are being processed by Level 2 processors, they are also 
stored on one of the four Level 2 buffers, waiting for 
the Level~2 trigger decision. Each buffer is emptied 
when the Level 2 decision for the corresponding event 
has been asserted: if the event has been accepted, the
buffer is read out, else it is simply cleared. 
If the Level~2 trigger decision takes 
too much time and the four buffers are all filled, 
the Level~1 accept is inhibited. This is a source 
of deadtime for the CDF~II trigger. The maximum 
Level 2 output rate is 300~Hz. The Level 3 trigger
is made of a CPU farm and has a maximum output rate
of 100~Hz.

The heart of the Level 1 trigger is the eXtremely
Fast Tracker (XFT, \cite{XFT} \cite{xft3d}), the trigger track
processor that identifies high transverse momentum
($p_T> 1.5$~GeV/c) charged tracks in the COT. 
The XFT tracks are three-dimensional and are 
extrapolated to the calorimeter
and to the muon chambers to generate electron and muon
trigger candidates. Track identification in the XFT
is accomplished in two processes by the Finders and 
by the Linkers. The Finders search for track segments 
in the axial and stereo superlayers of the chamber. 
The Linkers search for matches among segments in 
the superlayers, consistent with prompt high-$p_T$ tracks.
The efficiency for finding XFT tracks is 
$\approx$90~\%, with a transverse momentum resolution better
than 2~\% per GeV/c and azimuthal angular resolution
of 5.5~mr. The level of fake tracks shows some growth 
with instantaneous luminosity.

The Online Silicon Vertex Tracker (SVT, \cite{SVT}) is part
of the Level 2 trigger. It receives the list of XFT tracks
and the digitised pulse heights on the axial layers of the
silicon vertex detector. The SVT links the XFT tracks to the 
silicon hits and reconstructs tracks with offline-like quality.
In particular the resolution on the impact parameter, which is
a crucial parameter to select B events since they tipically 
show secondary vertices, is 35~$\mu$m for 2~GeV/c tracks. 
The SVT efficiency is ~85~\% per track.
Since a long Level 2 processing time can introduce dead time,
to speed up operations the SVT has a widely parallelized design: 
it is made of 12 identical azimuthal slices working in parallel.
Each slice receives and processes data only from one silicon
vertex detector 30$^{\circ}$ sector. In addition SVT recontructs
only tracks in the transverse plane to the beamline and only with
$p_T >$~2.0~GeV/c. The tracking process is performed in two steps.
The first step is the pattern recognition: candidate tracks are
searched among a list of precalculated low resolution patterns.
This is done in order to reduce the huge amount of silicon hits
only to those potentially interesting. The second step
is track fitting: a full resolution fit of the hit coordinates
found within each pattern is performed using a linearized algorithm.
By providing a precision measurement of the impact parameter 
of the charged particle tracks, SVT allows triggering on events
cointaining long lived particles, like the B events, which at 
the Tevatron have decay lengths of the order of 500~$\mu$m 
and produce tracks in the decay with impact parameters on
average larger than 100~$\mu$m.

Level 3 trigger is implemented on a CPU farm which allows 
to perform an almost offline-quality event reconstruction.

\subsection{Triggers for B physics}
CDF~II has basically three families of triggers for B physics:
the dimuon trigger, the semileptonic trigger and the hadronic 
trigger. 

The dimuon trigger selects muon pairs with
transverse momenum as low as 1.5~GeV/c. It is mostly used to
select $J/\psi$s and $\psi(2S)$, to reconstruct the many decay
modes of the B hadrons ($B^0$, $B^+$, $B^0_s$, $B_c$, and $\Lambda_b$)
containing a $J/\psi$ decaying to muon pairs, and to select 
$\Upsilon\rightarrow \mu^+\mu^-$ decays, or muon 
pairs for the search of the rare $B\rightarrow \mu^+\mu^- X$ decays,
or for $b\overline{b}$ correlation studies.

The semileptonic trigger selects events with a lepton ($\mu$ or $e$)
with $p_T~>$~4 GeV/c and an SVT track with $p_T >$~2~GeV/c and impact
parameter above 120~$\mu$m.

The hadronic trigger selects hadronic decay modes as
$B^0_{(s)}\rightarrow h^+ h^{'-}$ and $B^0_{s}\rightarrow D^-_s\pi^+$.
Te trigger requires track pairs with $p_T > 2$~GeV/c
and $p_{T1}+p_{T2} >$5.5~GeV/c, with an opening angle in the
transverse plane below 135$^{\circ}$,
impact parameter above 100~$\mu$m,
a decay length above 200~$\mu$m. For the two-body decay
trigger path, optimised to collect $B\rightarrow h^+h^{'-}$ decays, 
the track pair is requested to point back to the
primary vertex, by requiring that the impact parameter of the 
reconstructed B is below 140~$\mu$m. To select hadronic multibody
decays, like $B^0_s\rightarrow D^-_s\pi^+$, the request on the
pointing back to the primary vertex has low efficiency, since the
track pair provides only a partial reconstruction of the multibody
decay, and it is not applied.

The Tevatron performance has been continuosly improving since the 
beginning of Run~II. The high initial luminosity and the drop along
the store require continuous monitoring and adjustments of the 
trigger strategy in order to have all the available bandwidth
efficiently used at all luminosities. This has been achieved by 
implementing more versions of the same trigger with increasingly 
tighter cuts, and consequently increasing purity, and 
running the versions with tighter cuts with no prescale and 
the versions with looser cuts with a prescale. For the hadronic
trigger the tighter version requires track pairs with $p_T >$ 2.5 GeV/c
and $p_{T1}+p_{T2} >$ 6.5 GeV/c. The prescales can 
be made dynamical, and they adjust themselves as the luminosity 
decreases and more bandwidth becomes available. In addition to this,
triggers are luminosity enabled: if the rate of a trigger that is
important to keep unprescaled is too large
at the initial luminosity, it is enabled only when the luminosity has
decreased below a safe threshold.

\section{First Bounds on Mixing-Induced CP Violation
in $B^0_s\rightarrow J/\psi\phi$ Decays}
\begin{figure}
 \includegraphics[height=9.0cm]{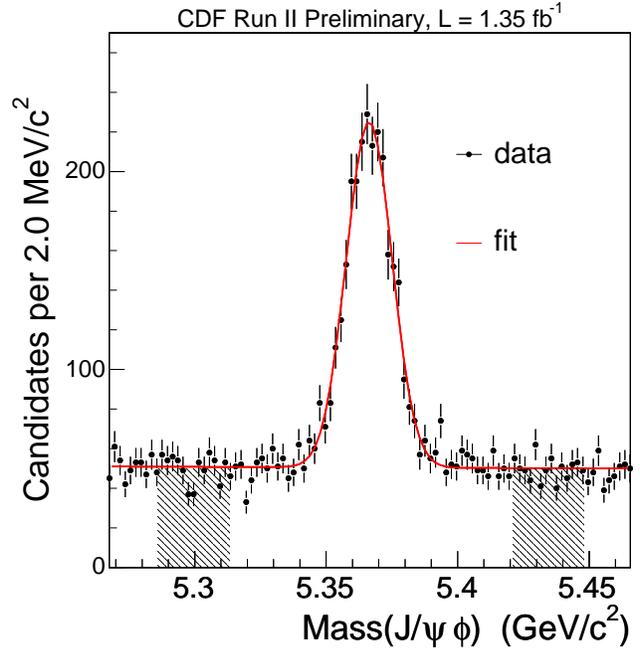}%
 \caption{invariant $\mu^+\mu^- K^+ K^-$ mass distribution with 
the fit projection overlaid.\label{fig:B0smass}}
 \end{figure}

\begin{figure}
\centering
 \includegraphics[height=8.2cm]{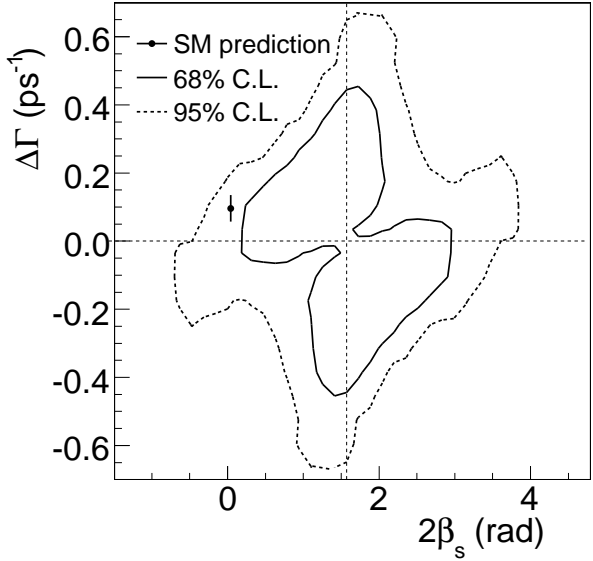}%
 \caption{Feldman-Cousins confidence region in the
$2\beta_s - \Delta\Gamma$ plane, where the standard model 
favored point is shown with error bars. The intersection
of the horizontal and vertical dotted lines indicates the reflection
symmetry in the $2\beta_s$ - $\Delta\Gamma$ plane.\label{fig:bs_dgamma}}
 \end{figure}

CDF has performed a first study of the $B^0_s\rightarrow J/\psi\phi$ 
decay in which the initial state is identified as $B^0_s$, or its
antiparticle $\overline{B}^0_s$ by using the process of flavor tagging.
This information is used to separate the time evolution of mesons 
produced as $B^0_s$ or $\overline{B}^0_s$. By relating this time
development with the CP eigenvalue of the final state that is 
accessible through the angular distribution of the $J/\psi$ and
$\phi$ mesons, we obtain direct sensitivity to the CP-violating
phase. This phase enters the time development with terms 
proportional to both $\mid \cos(2\beta_s)\mid$ and $\sin(2\beta_s)$.
Analyses of $B^0_s\rightarrow J/\psi\phi$ decays that do not use
flavor tagging provide information on $\Delta\Gamma$, and are 
primarily sensitive to $\mid \cos(2\beta_s)\mid$ and 
$\mid\sin(2\beta_s)\mid$, leading to a fourfold ambiguitiy in 
the determination of $2\beta_s$ \cite{betas1}\cite{betas2}.
This analysis uses 1.35 fb$^{-1}$ of data collected by the dimuon 
trigger, which selects events containing $J/\psi\rightarrow\mu^+\mu^-$
decays. The $B^0_s\rightarrow J/\psi\phi$ decays are 
reconstructed from the decays  $J/\psi\rightarrow\mu^+\mu^-$
and $\phi\rightarrow K^+ K^-$ and require these final states 
to originate from a common vertex. An artificial neural network
is used to achieve optimal separation of signal from background
\cite{ANN}.
The neural network uses particle identification information from
the time-of-flight detector and the specific ionization loss in
the drift chamber, the transverse momentum components of the
$B^0_s$ and $\phi$ candidates and the quality of the fit to the
trajectories of the final state particles. For the network training
MonteCarlo events are used as signal, while background events are
from the $B^0_s$ mass sidebands in real data.
The analysis uses the orbital angular momenta of the vector mesons
produced in the decay of a pseudoscalr meson to distinguish 
between the CP-even (S- and D-wave) from the CP-odd (P-wave) 
final states. To separate the time development of the B$^0_s$ 
and $\overline{B}^0_s$, the flavor at the time of production is
identified by means of two independent types of flavor tagging.
The first type, known as opposite-side flavor tag, uses the decay
products of the other $b$ quark in the event, and is mainly based
on the charge of the muons or electrons form the semileptonic B decays
\cite{OST}.
The second type correlates the flavor of the meson with the charge
of an associated kaon, referred as same-side kaon, 
arising from fragmentation processes \cite{SST}. The parameters of interest,
$2\beta_s$ and $\Delta\Gamma$, plus several additional parameters,
including the mean $B^0_s$ width $\Gamma$, the mixing frequency $\Delta m_s$
and the magnitudes of the polarisation amplitudes and the strong phases,
are extracted from an unbinned maximum likelihood fit.
The fit uses the $B^0_s$ candidate mass $m$ and its undertainty
$\sigma_m$ (Figure \ref{fig:B0smass}), 
the $B^0_s$ proper decay time $t$ and its uncertainty
$\sigma_t$, the transversity angles \cite{betas1}, 
and the tag information.
A confidence region in the $2\beta_s$-$\Delta\Gamma$ plane is
constructed using the Feldman-Cousins methods \cite{FC}, reported in 
Figure \ref{fig:bs_dgamma}.
Assuming the standard model predicted values of 
$2\beta_s = 0.04$ and $\Delta\Gamma = 0.096$~ps$^{-1}$,
the probability of a deviation as large as the level observed in 
data is 15~\% (1.5$\sigma$). If $\Delta\Gamma$ is treated as a
nuisance parameter, and only $2\beta_s$ is determined from the
fit, $2\beta_s$ falls in the window [0.32,2.82] at the 68~\%
confidence level. The new bounds restrict the knowledge of 
$2\beta_s$ to two of the four solutions allowed in the measurements
that do not use flavor tagging \cite{betas1}\cite{betas2}.

\section{Search for the $B^0_{(s)}\rightarrow \mu^+\mu^-$ rare Decay}
\begin{figure}
 \includegraphics[height=8.0cm]{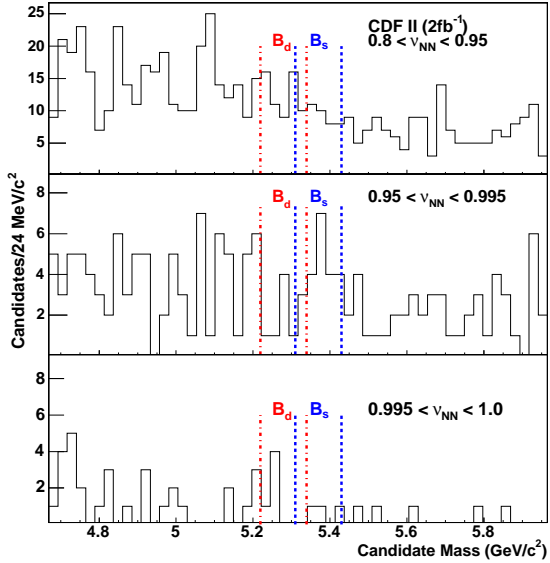}%
 \caption{The $\mu^+\mu^-$ invariant mass distribution
for events satisfying all selection criteria for the
final three ranges of the neural network variable.
\label{fig:bsmumu}}
 \end{figure}

The FCNC decays $B^0_s(B^0)\rightarrow\mu^+\mu^-$ occur in the
standard model only through higher order diagrams. The expectations
for the branching fractions are
$BR(B^0_s\rightarrow\mu^+\mu^-) = (2.42\pm0.54)\times 10^{-9}$
and 
$BR(B^0\rightarrow\mu^+\mu^-) = (1.0\pm0.14)\times 10^{-10}$
\cite{bsmumubr} which are one order of magnitude smaller than
the current experimental sensitivity. In SUSY models diagrams
including sypersymmetric particles can increase these branching
fractions at large $tan(\beta)$ \cite{bsmumuth1}, and also at small
$tan(\beta)$ \cite{bsmumuth2}.
The new measurement uses 2 fb$^{-2}$ and the sensitivity is further
improved by using an enhanced muon selection, and performing the 
search in a two dimensional grid in dimuon mass and neural network
space. The discriminating variables used in the analysis include
the measured proper decay time, $\lambda$, the proper decay time
divided by the estimated uncertainty, $\lambda/\sigma_{\lambda}$,
the 3D opening angle between the dimuon momentun and the displacement
vector between the primary vertex and the dimuon vertex,
$\Delta\Theta$, and the B-candidate track isolation, $I(B)$.
To enhance signal and background separation a neural network
discriminant is constructed, using all the discriminating
variables except $m_{\mu\mu}$. The network is trained using
MonteCarlo simulated $B^0_{(s)}$ events for signal and 
mass sidebands events for background.
The mass search regions are 
$5.310 < m_{\mu\mu} < 5.430$~GeV/c$^2$ for the $B^0_s$
and
$5.219 < m_{\mu\mu} < 5.339$~GeV/c$^2$ for the $B^0$,
which approximately correspond to $\pm2.5\sigma_m$ 
($\sigma_m = 24$~MeV/c$^2$).
The $B^0_s\rightarrow\mu^+\mu^-$ branching fraction 
is determined relative to the $B^+\rightarrow J/\psi K^+$
signal. Relative geometric acceptances and analysis 
efficiencies are estimated using MonteCarlo.
The expected background is due to the contributions 
of the combinatorial continuum and from the 
$B^0_{(s)}\rightarrow h^+ h^{'-}$ decays.The contribution
from the combinatorial is estimated by linearly extrapolating
from the sideband region to the signal region. The contribution
from $B^0_{(s)}\rightarrow h^+ h^{'-}$ decays is estimated to
be one order of magnitude smaller than the conbinatorial
background. Using an a priori optimization procedure, it 
was found that subdividing the signal region into several
bins in mass and in the neural network variable improves
the sensitivity of the analysis relative to using a single
bin. The backgrounds, efficiencies and limits are computed
in each bin separately. Figure \ref{fig:bsmumu} shows the
$\mu^+\mu^-$ mass distribution for three different ranges
of the neural network variable.
We extract the following limits
$BR(B^0_s\rightarrow\mu^+\mu^-)< 5.8(4.7)\times 10^{-8}$
at 95(90)~\% C.L. and
$BR(B^0\rightarrow\mu^+\mu^-)< 1.8(1.5)\times 10^{-8}$
at 95(90)~\% C.L., which place further constraints on 
new-physics models \cite{bsmumuth1}\cite{bsmumuth2}.

\section{New $B_c$ and $B^0_s$ Lifetime Measurements}
 \begin{figure}
 \includegraphics[height=8.0cm]{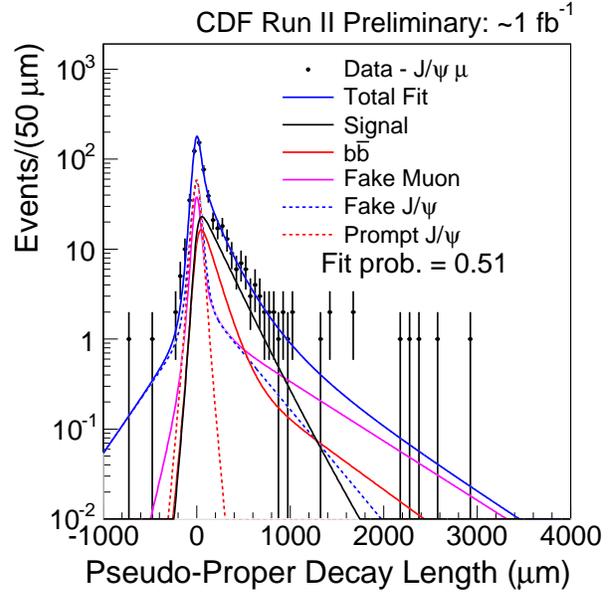}%
 \caption{Fitted $ct^{*}$ for the $J/\psi \mu$ candidate 
events, where background are shown broken into individual 
components.
\label{fig:bctaumu}}
 \end{figure}

\begin{figure}
 \includegraphics[height=8.0cm]{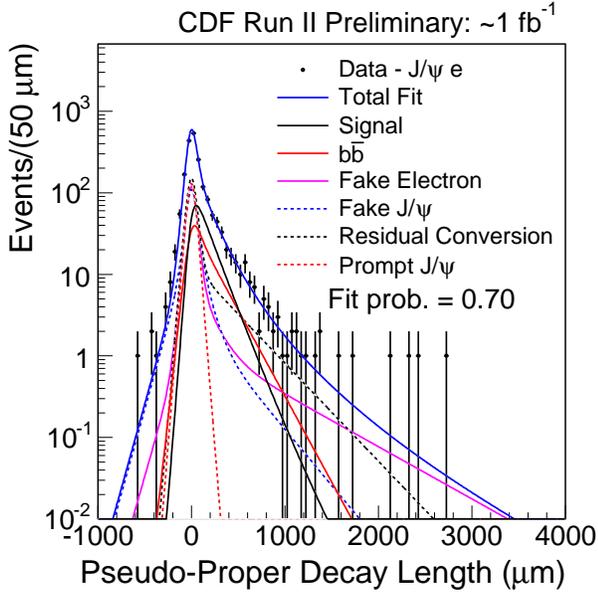}%
 \caption{Fitted $ct^{*}$ for the $J/\psi e$ candidate 
events, where background are shown broken into individual 
components.
\label{fig:bctaue}}
 \end{figure}

CDF has measured the $B^+_c$ lifetime in 1 fb$^{-1}$ of inclusive
$J/\psi\rightarrow\mu^+\mu^-$ data using the
$B^{\pm}\rightarrow J/\psi l^{\pm}X$ decays, where $l$ can be an
electron or a muon and $X$ are unmeasured particles.
To measure the lifetime, we reconstruct a per event lifetime defined
using variables measured in the transverse plane. Since not all the
decay products of the $B_c$ decay are identified, we define a
\begin{equation}
ct^{*} = \frac{mL_{xy}(J/\psi l)}{p_T(J/\psi l)}
\end{equation}
from which we can obtain the true $B_c$ lifetime by defining
a factor $K$, determined using MonteCarlo, where $ct = Kct^{*}$.
The $B_c$ signal is reconstructed by selecting dimuon events 
with a di-muon mass falling within $\pm50$~MeV/c$^2$ window 
around the $J/\psi$ mass, and requiring the presence of a third 
lepton forming a vertex with the $J/\psi$ candidate. Most 
$B_c$ signal is expected in the window with
$4.0 < m_{J/\psi l} < 6.0$~GeV/c$^2$.
The broad mass peak includes also significant amount of 
background, mainly due to events with a $J/\psi$ from $b\overline{b}$
and $c\overline{c}$ and an additional hadron which fakes a muon by
punching through to the muon chambers, or an electron by leaving a
signature in the calorimeter compatible with an electron, to events
from the continuum background for $J/\psi$ with a third lepton,
to $b\overline{b}$ events with a $J/\psi$ froma $b$-quark jet
and a lepton from the other b, or events where conversions produce
electrons which produce a $B_c$ candidate when paired to a $J/\psi$.
Each of these background sources are estimated by using control samples
and simulation. The fit of the $ct^{*}$ distributions is performed
separately in the $J/\psi l$ and $J/\psi e$ decay modes, using
likelihood functions (Figure \ref{fig:bctaumu} and \ref{fig:bctaue}).
The combined result is 
$c\tau = 142.5^{+15.8}_{-14.8} (stat.)\pm 5.5 (syst.)$~$\mu$m, where
the largest contributions to the systematic error is due to the 
resolution function and to the uncertainty on the silicon alignment.

\begin{figure}
 \includegraphics[height=11.cm]{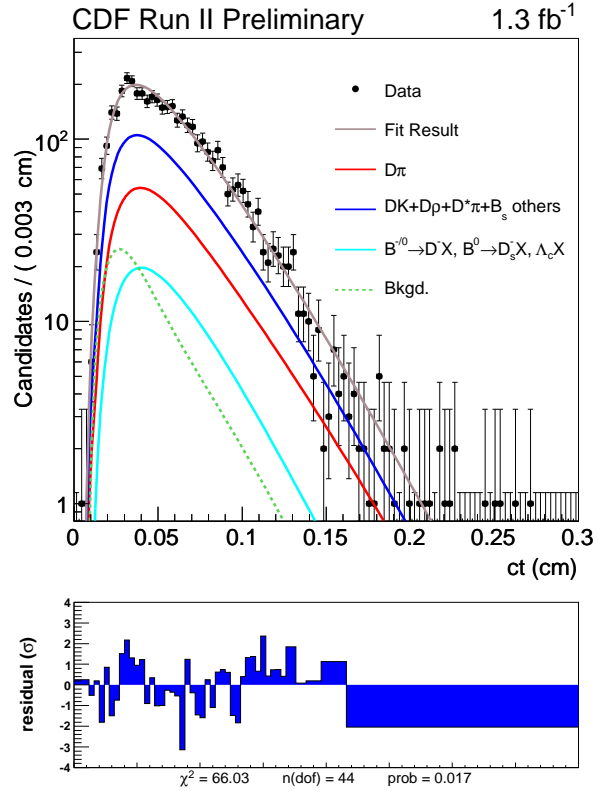}%
 \caption{$ct$ projection of the lifetime fit for events
reconstructed as $B^0_s\rightarrow D^-_s(\phi\pi^-)\pi^+$.
\label{fig:bstauh}}
 \end{figure}

A new CDF analysis has measured the $B^0_s$ lifetime using
1.3~fb$^{-1}$ of data reconstructed as 
$B^0_s\rightarrow D^-_s(\phi\pi^-)\pi^+$ and
$B^0_s\rightarrow D^-_s(\phi\pi^-)\rho^+(\pi^+\pi^0)$, 
where the $\pi^0$ is not reconstructed.
The analysis reconstructs about 1100 fully reconstructed
$B^0_s\rightarrow D^-_s(\phi\pi^-)\pi^+$  candidates and a 
similar number of partially reconstructed
$B^0_s\rightarrow D^-_s(\phi\pi^-)\rho^+(\pi^+\pi^0)$ decays.
The increased statistics has an uncertainty due to the missing
tracks or misassigned masses, but it can be properly accounted 
for and folded into the likelihood formulation. The lifetime
fit of the $B^0_s$ meson is determined from two fits done
sequentially. The first fit is of the reconstructed mass
of the $B^0_s$ and it determines the relative fractions
of the various decay modes and backgrounds. The second 
fit is of the proper decay time of the $B^0_s$ candidates
and uses the fractions determined with the mass fit
(Figure \ref{fig:bstauh}).
The analisys determines
$c\tau = 455.0\pm12.2(stat.)\pm 7.4(syst.)$, where the
main systematics are due to the background model and to
the uncertainty in the measurement of the fractions of
the several contributions, to the uncertainty of the 
shape of the MonteCarlo $p_T(B^0_s)$ distribution and
to the silicon detectors alignment.

\section{CP Violation in the $B\rightarrow DK$ and 
$\Lambda^0_b\rightarrow h^+h^{'-}$ decays}
CDF has also performed analyses in the field of CP violation
using hadronic decays reconstructed in 1 fb$^{-1}$ of data. 
A first analysis uses the 
$B^-\rightarrow D^0 K^-$ modes, which allow a theoretically-clean
extraction of the CKM angle $\gamma$ by a variety of methods, depending 
on the involved specific $D^0$ decay mode \cite{BDKth1}\cite{BDKth2}
\cite{BDKth3}\cite{BDKth4}. These methods require no
flavor tagging or time-dependent measurements and use only charged
particles in the final state, which makes them well suited for a
hadron collider environment.
By using the data collected by the hadronic trigger we have reconstructed
the modes where the $D^0$ decays to either $K^-\pi^+$ (flavor eigenstate)
or $K^-K^+$, $\pi^-\pi^+$ (CP-even eigenstate). The relevant modes are
separated from the larger $B^-\rightarrow D^0\pi^-$ modes by using a
likelihood fit which exploits the kinematic differences among the decay
modes and the particle identification information provided by the 
specific ionization in the drift chamber.
The most relevant result of this analysis is the first measurement at
a hadron collider of the direct CP asymmetry in the 
$B^-\rightarrow D^0_{CP+} K^-$ decay mode,
$A_{CP+} = 0.37\pm0.14(stat.)\pm0.04(syst.)$, which is in good 
agreement with the measurements performed at the $\Upsilon(4S)$
and has a comparable resolution. The analysis measures also other
decay parameters useful to improve the determination of the CKM
angle $\gamma$.

\begin{figure}
 \includegraphics[height=8.cm]{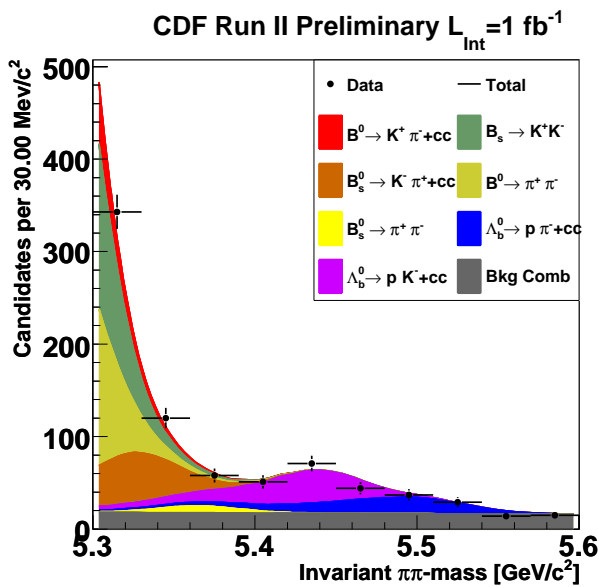}%
 \caption{Invariant di-hadron mass distribution in the
$\Lambda^0_b$ mass window with the result of the fit overlaid.
\label{fig:Lbhh}}
 \end{figure}

CDF has also performed a first measurement of CP-violating
asymmetries and branching fractions of $\Lambda^0_b$ charmless
two-body decays. These decays in a proton and a charged kaon
or pion may show significant CP-violating asymmetries, of the
order of 30~\%. A measurement of these asymmetries may be useful
to allow or rule out some extensions of the standard model
\cite{lbhhth1}\cite{lbhhth2}.
The analysis uses the same general framework used for the
measurement of two-body charmless decays of the B mesons,
reported in \cite{b0hhprl}. An unbinned maximum likelihood 
fit, using kinematic information as well as particle identification
information from the specific ionization measurement in the drift
chamber, is used to measure the different components of the
$\Lambda^0_b$ signal. These variables allow also to separate
the $\Lambda^0_b$ from the $\overline{\Lambda}^0_b$ decays,
thus allowing for a measurement of the direct CP asymmetry.
The necessary corrections which take into account detector
acceptance effects and trigger and offline selection efficiencies
are determined from control samples and from simulated samples.
Figure \ref{fig:Lbhh} reports the di-hadron mass distribution 
reconstructed in data in the $\Lambda^0_b$ window. 
We have measured $A_{CP}(\Lambda^0_b\rightarrow p\pi^-)
= 0.03\pm0.17(stat.)\pm 0.05(syst.)$ and
$A_{CP}(\Lambda^0_b\rightarrow pK^-)
= 0.37\pm0.17(stat.)\pm 0.03(syst.)$

\section{Evidence for $D^0$ Mixing and Search for the 
$D^0\rightarrow\mu^+\mu^-$ rare Decay}
\begin{figure}
\vspace*{0.5cm}
 \includegraphics[height=7.cm,width=8.cm]{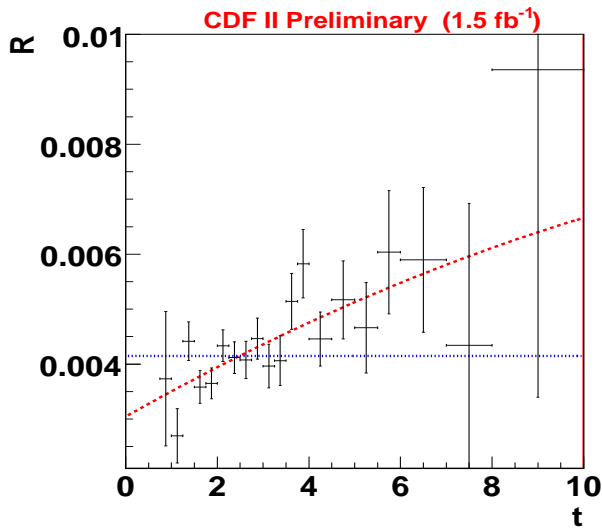}%
 \caption{Ratio of the prompt $D^*$ ``wrong-sign'' to ``right-sign''
decays as a function of normalized proper decay time. The dashed curve
is from a least-squares quadratic fit. The mixing parameters are 
determined from this fit. The dotted line is the best fit if we assume
no mixing.
\label{fig:dmix1}}
 \end{figure}

\begin{figure}
\hspace*{-0.8cm}
 \includegraphics[height=8.cm,width=9.cm]{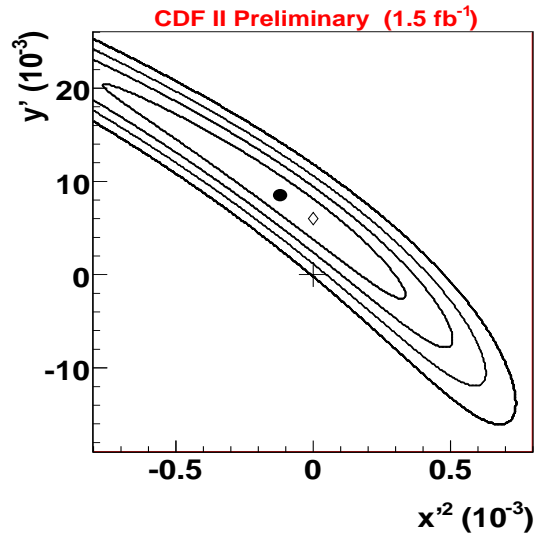}%
 \caption{Bayesian probability countours in the parameter space
corresponding to one through four equivalent Guassian standard
deviations. The closed circle is the best fit value for the 
mixing parameters. The open diamond is the highest probability
point that is physically allowed. The cross is the no-mixing point.
\label{fig:dmix2}}
 \end{figure}

The recent evidence for $D^0$ mixing has been found in two different
types of measurements. The BELLE Collaboration found direct evidence
of a longer and shorter lived $D^0$ meson \cite{d0mix_belle}, the
BaBar Collaboration found a difference in decay time distribution
for $D^0\rightarrow K^+\pi^-$ compared to that for 
$D^0\rightarrow K^-\pi^+$ \cite{d0mix_babar}. CDF has performed
a measurement comparing the decay time distribution for
$D^0\rightarrow K^+\pi^-$ compared to that for
$D^0\rightarrow K^-\pi^+$. The ratio $R(t)$ of $K^+\pi^-$ to
$K^-\pi^+$ can be expressed as a simple quadratic function of
proper time $t$ under the assumption of CP conservation and
small values for the parameters $x$ and $y$ \cite{d0mix_th1},
where the parameter $x$ is defined in terms of the mass
difference $\Delta m$ between the heavy and light mass eigenstates
and the parameter $y$ involves the width difference $\Delta\Gamma$
between these states. The CDF measurement uses 1.5~fb$^{-1}$ of data
collected by the hadronic trigger. We reconstruct the Cabibbo-favored
decay chain $D^{*+}\rightarrow D^0\pi^+$, with $D^0\rightarrow K^-\pi^+$
called ``right sign'' and the corresponding ``wrong sign'' and the 
selection cuts are chosen as to maximise the significance of the
wrong sign signal. The ratio of wrong sign to right sign $D^{*}$
is determined by dividing the time distribution in bins. The distribution
is fit to a parabola, related to the mixing parameters
(Figure \ref{fig:dmix1}).
The uncertainty on the mixing parameters includes effects from
statistical fluctuations, uncertainties from the signal and
background shapes and the corrections due to charm mesons from
B decays. A Bayesian method is used to get the probability for 
different mixing parameters values. Contours containing the highest 
probability points are shown in Figure \ref{fig:dmix2}. 
The no-mixing point lies outside the contour equivalent 
to $3.8\sigma$ standard deviations, which is evidence
of chamr mixing.

CDF has also searched for the rare $D^0\rightarrow \mu^+\mu^-$
decay, which is suppressed by the GIM mechanism in the standard
model which predicts a branching fraction of the order of
$10^{-13}$ \cite{d0mumuth}, with possible enhancements of
seven orders of magnitude in R-parity violating
SUSY models. The CDF anaysis has been performed with 360 pb$^{-1}$
of hadronic data and follows the same strategy as in the previous 
CDF measurement \cite{d0mumuprd}. The two-track candidates are 
reconstructed with the $\mu^+\mu^-$ mass hypothesis, and the 
$D^{*-}$ tag mass-difference is applied to reduce the combinatorial
background. By using the well known $D^0\rightarrow\pi^+\pi^-$ decay
mode as a normalisation mode, the analysis set the limit
$BR(D^0\rightarrow\mu^+\mu^-) < 0.43(0.53)\times 10^{-6}$
at the 90(95)~\% C.L.

\section{Conclusions}
In this paper we have reviewed the current CDF hot topics in the 
fields of $b$ and $c$ physics. Most current CDF physics results 
are producd with 1-2 fb$^{-1}$. With the already collected 
3.5~fb$^{-1}$ and the total 6~fb$^{-1}$ expected by the end
of the year 2009, CDF expects to increase signficantly the 
number and the quality of the physics results in these sectors.

\bigskip 

\end{document}